\documentclass[fleqn,10pt]{wlscirep}
\usepackage[utf8]{inputenc}
\usepackage[T1]{fontenc}
\usepackage{url}
\usepackage{hyperref}
\usepackage[numbers]{natbib}

\title{Matters Arising on ``Breakup of a long-period comet as the origin of the dinosaur extinction" by Siraj \& Loeb}

\author[1,*]{Steven J. Desch}
\author[1]{Alan P. Jackson}
\author[2]{Jessica L. Noviello}
\author[1,3]{Ariel D. Anbar}

\affil[1]{School of Earth and Space Exploration, Arizona State University, Tempe AZ 85287-1404}
\affil[2]{NASA Postdoctoral Management Fellow, NASA Goddard Space Flight Center, Greenbelt, MD 20771}
\affil[3]{School of Molecular Sciences, Arizona State University, Tempe, AZ 85287-1604}

\affil[*]{steve.desch@asu.edu}

\affil[+]{these authors contributed equally to this work}

\begin{abstract}
The recent publication by Siraj \& Loeb (2021; {\it Nature Scientific Reports} 11, 3803) attempts to revive the debate over whether the Chicxulub impactor was a comet or an asteroid. 
They calculate that $\sim$20\% of long-period comets impacting Earth will have first been disrupted by passage inside the Sun's Roche limit, generating thousands of fragments, each the needed size of the Chicxulub impactor. 
This would increase the impact rate of comets by a factor $\sim 15$, making them as likely to hit the Earth as an asteroid.
They also argue that a comet would be a factor of 10 more likely to match the geochemical constraints, which indicate the Chicxulub impactor was carbonaceous chondrite-like. 
These conclusions are based on misinterpretations of the literature.
Siraj \& Loeb \citep{Siraj+Loeb2021} overestimate the number of fragments produced during tidal disruption of a comet: tens of fragments are produced, not thousands.
They also conflate 'carbonaceous chondrite' with specific types of carbonaceous chondrite, and ignore the evidence of iridium, making comets seem more likely than asteroids to match the Chicxulub impactor, when in fact they likely can be ruled out.
Rather than a comet, an asteroidal impactor similar to CM or CR carbonaceous chondrites is strongly favored. 
\end{abstract}
\begin{document}

\flushbottom
\maketitle
\thispagestyle{empty}

\section*{Introduction}

Since the discovery of Ir in the clay layer at the K-Pg boundary \citep{Alvarez+1980}, scientists have sought to constrain the origin of the extraterrestrial impactor that triggered the end-Cretaceous mass extinction of the non-avian dinosaurs and other species.
While the first proposal was for an asteroid \citep{Alvarez+1980}, for a while some theories invoked a cometary impactor to explain perceived periodicities in mass extinctions \citep{Rampino+Stothers1984}.
Such models have long been disfavored by the mass of Ir in the layer, inferred to be $2.0 - 2.8 \times 10^{11} \, {\rm g}$ \citep{Artemieva+Morgan2009}.
The size of Chicxulub crater leads to an estimated asteroid impactor diameter, $D \approx 10 \, {\rm km}$ \citep{Brittan1997, Ivanov2005}. For a cometary impactor this decreases to $D \approx 7 \, {\rm km}$ due to the higher impact speed \citep{Brittan1997}. 
A carbonaceous chondrite-like asteroid of this size would likely deliver $\approx 2.3 \times 10^{11} \, {\rm g}$ or Ir \citep{Brittan1997}, in the center of the estimated mass range of the global Ir layer; but a comet would only deliver $\sim 0.1 \times 10^{11} \, {\rm g}$, because it is smaller and mostly ice.
Against this backdrop, Siraj \& Loeb \citep{Siraj+Loeb2021} have argued in favor of a comet over an asteroid, based on dynamical and geochemical evidence.  
Here we demonstrate that their arguments are based on misinterpretations of the literature, and that an asteroid is in fact highly favored over a comet.

\section*{Geochemical Arguments}

Siraj \& Loeb \citep{Siraj+Loeb2021} cite good evidence that the Chicxulub impactor was carbonaceous chondrite-like, but then assert that 100\% of comets satisfy this constraint but only 10\% of asteroids do. 
This assertion conflates carbonaceous chondrites with specific types (CB, CH, CI, CM, CO, CR, CV) of carbonaceous chondrites.
It underestimates the fraction of asteroids that match the Chicxulub impactor's composition, and/or overestimates the fraction of comets that would.

Siraj \& Loeb \citep{Siraj+Loeb2021}, citing Bottke et al.\ \citep{Bottke+2007}, claim only 30\% of asteroids are C-type (spectrally resembling carbonaceous chondrites) and appear to imply that only 40\% of carbonaceous chondrites are the specific type CM associated with the impactor.
In fact the fraction of asteroids that are C-type is $> 50\%$ \citep{Morbidelli+2020}. 
As well, the Chicxulub impactor could be CM- {\it or} CR-like. 
Siraj \& Loeb \citep{Siraj+Loeb2021} cite evidence from a fossil meteorite in the K-Pg clay layer, which demands the impactor be CV, CO, CR, or possibly CM, but not CI \citep{Kyte1998}. 
They also cite evidence from the $\epsilon^{54}{\rm Cr}$ isotopic anomaly in the K-Pg clay clayer, which argues the impactor was CM (and CR, CH, and CB have the same $\epsilon^{54}{\rm Cr}$), but argues against CV, CO, and CI \citep{Trinquier+2006}.
The authors could have cited equally strong arguments from platinum-group element patterns, which favor CM or CO (and allow CR), but rule out CI \citep{Goderis+2013}.
The composition of the Chicxulub impactor is a match to either CM or CR chondrites.
Siraj \& Loeb \citep{Siraj+Loeb2021} argue that CM chondrites comprise a fraction $\approx  40\%$ of all carbonaceous chondrites, based on statistics of intact falls; but %in fact the vast majority of meteoritic material falling to Earth arrives as micrometeorites, which are overwhelmingly CM- and CR-like \citep{Engrand+Maurette1998}. Therefore 
a larger fraction of C-type asteroids may match CM or CR chondrites. 
At a minimum, $\approx 50\%$ of asteroids are carbonaceous chondrite-like, and $>20\%$ of asteroids striking Earth match the specific composition of the Chicxulub impactor.

Siraj \& Loeb \citep{Siraj+Loeb2021} claim 100\% of comets are carbonaceous chondrite-like, which may be loosely true; but comets are not definitively associated with any particular subtype of carbonaceous chondrite, but are most strongly  associated  with carbonaceous chondrites of type CI, based on their low albedo, friability, lack of chondrules, presence of anhydrous silicates, and low impact rate on Earth \citep{Campins+Swindle1998}. 
A comet-like origin has been argued for CI chondrites like Orgeuil \citep{Gounelle+2006}. 
None of the lines of geochemical evidence above is consistent with CI chondrites, indicating that while 100\% of comets may be carbonaceous chondrite-like, possibly 0\% of them match the specific composition of the Chicxulub impactor in detail. 

Siraj \& Loeb \citep{Siraj+Loeb2021} applied a double standard to the geochemical evidence. 
If the impactor must only be carbonaceous chondrite-like, then comets are more likely (for a given impact rate) by a factor of 2, not 10. 
If the impactor must match a CM or CR composition, then $> 20\%$ of asteroids provide a match, but no comets do. 
The mass of Ir in the clay layer likewise is a match to an asteroidal impactor, but not a comet. 

\section*{Dynamical Arguments}

Siraj \& Loeb \citep{Siraj+Loeb2021} downplay the frequency with which asteroids impact Earth, and overestimate the likelihood of a comet impact. 
The authors state that the Chicxulub impact was the single largest impact in the last 250 Myr, and that asteroids with $D= 10$ km should impact the Earth with mean rate once per $\sim 350$ Myr. Therefore by their own numbers the probability of a $D > 10$ km asteroid impacting Earth in the last 250 Myr is $> 50\%$. Whatever the probability of a comet impact, an asteroid impactor is a probable event. 

The main point of Siraj \& Loeb \citep{Siraj+Loeb2021} is that a significant fraction, $\sim 20\%$, of long-period comets (LPCs) impacting the Earth will have first passed through the Sun's Roche limit and fragmented into a number, $N$, of smaller comets, potentially increasing the probability one will strike Earth. 
A comet $N$ times more massive than the final Chicxulub impactor is rarer than an undisrupted LPC with the size of the Chicxulub impactor, by a factor of $(N^{1/3})^{1-q}$, where $q \approx 2.0 - 2.7$; but because there are more fragments, this would increase the rate of Chicxulub-scale impactors by a factor $\approx 0.2 \, \times \, N \, \times \, N^{(1-q)/3}$, which is $\approx 15$ for $q = 2$ and $N = 630$ (equivalent to a 60 km-diameter comet breaking up into ones with diameter 7 km). 
The authors state that undisrupted LPCs the size of the Chicxulub impactor ($D = 7$ km) are expected to impact Earth once every 3.8 - 11 Gyr, so only if $N \sim 10^3$, enhancing the fluxes by factors $> 15$, is the collision timescale $\approx 250$ Myr and comparable to asteroids. 

Despite its importance, the choice of $N \approx 630$ appears unjustified.
Tidal disruption of comets like Shoemaker-Levy 9, and crater chains on Jupiter's moons, suggest a value closer to $\approx 20$. 
The analytical treatment of Hahn \& Rettig \citep{Hahn+Rettig1998} shows the number of fragments generated is fixed during the encounter, by the relative timescales of spreading and gravitational contraction, which are functions of the comet's density, $\rho_0$, and its perihelion distance, $r_0$. 
The contraction timescale, $t_{\rm contr}$, in units of the encounter timescale, $\tau =(G \rho_{\rm c})^{-1/2}$, is $t_{\rm contr} / \tau \approx 0.94 \, (\rho_{\rm c}/ \rho_0)^{1/2} \, N^{1/2}$, where $\rho_{\rm c} = (1 M_{\odot}) / r_0^3$.
The spreading timescale in units of the encounter timescale is found by numerical simulation and appears to be $t_{\rm spread} / \tau \approx 0.7 \, N^{0.85}$, assuming the dimensionless treatment applies to the Sun as well as Jupiter.  
A disrupted comet coalesces into fragments when these timescales are equal, which is when $N \approx 2.3 (\rho_{\rm c} / \rho_0)^{1.43}$. 
The closer to the Sun the comet penetrates, the more fragments are produced, but the minimum value of $r_0$, $1 R_{\odot}$, corresponds to $\rho_{\rm c} = 5.9 \, {\rm g} \, {\rm cm}^{-3}$. Assuming the authors' $\rho_0 = 0.7 \, {\rm g} \, {\rm cm}^{-3}$, the {\it maximum} number of fragments that can be produced by tidal disruption is $\sim 50$, for comets unrealistically skimming the Sun's photosphere. 
Assuming a more typical $r_0 \approx 0.7 \times$ the Roche limit, $N \approx 12$ is more likely.  That this is similar to the number of fragments produced in the tidal disruption of Shoemaker-Levy 9 is unsurprising since Jupiter and the Sun are of similar density.
The estimate $N \sim 10^{3}$ made by Siraj \& Loeb \citep{Siraj+Loeb2021} appears to be based on a misinterpretation of Hahn \& Rettig \citep{Hahn+Rettig1998}, somehow setting $N$ equal to $t / \tau$, where $t$ is the time for the fragments to reach Earth.

In addition, applying the formulism of Hahn \& Rettig \citep{Hahn+Rettig1998} to the case of a $D = 60$ km comet rounding the Sun, the length of the debris chain would be roughly 50 Earth diameters. Supposing $\sim 10^3$ fragments were generated and distributed over this length, the Earth would have collided with $\sim 20$ of them. This would lower the effective number of fragments to a maximum of $\sim 50$, leading to very little enhancement of the probability of a comet impact. It also would demand the Chicxulub impact be one of a chain of $\sim 20$ craters on Earth, which is not observed. 

\section*{Summary}

Siraj \& Loeb \citep{Siraj+Loeb2021} make a valid point that a Chicxulub-scale cometary impactor ($D = 7$ km) may be not {\it quite} as uncommon as previously thought, because some fraction of comets may be tidally disrupted by passage within the Sun's Roche limit. 
Similar ideas were expressed by Bailey et al.\ \citep{Bailey+1992}. 
But even setting $q=2$ and $r_0 = 1 R_{\odot}$, so that $N=50$, the enhancement in flux is only a factor $< 4$; and using the more likely $N=12$, the enhancement is only a factor of 2.
The mean timescale for an impact with a Chicxulub-scale comet is most likely $> 2$ Gyr, while the mean timescale with an asteroid remains $\sim 350$ Myr. 

Siraj \& Loeb \citep{Siraj+Loeb2021} applied a double standard to the geochemical evidence. 
If only a loose match to a carbonaceous chondrite is demanded, then comets are only a factor of 2, not 10, more likely than asteroids (for the same impact rate). 
If it is demanded that the impactors match a CM or CR carbonaceous chondrite composition, then $>20\%$ of asteroids, but possibly $\sim 0\%$ of comets, are a match. 
As well, Siraj \& Loeb \citep{Siraj+Loeb2021} cite Alvarez et al.\ \citep{Alvarez+1980} but ignore the evidence from the iridium in the K-Pg clay layer that is the point of that paper, which favors an asteroidal impactor but strongly disfavors a comet, which only supplies about 4\% as much iridium as an asteroid \citep{Brittan1997}. 

There is a $>50\%$ probability a $D=10$ km asteroid would have hit the Earth in the last 250 Myr. 
Among Earth-crossing asteroids, $\approx 50\%$ are C-type, associated with carbonaceous chondrites. 
At least $40\%$ of C-type asteroids, possibly more, will be of the type CM or CR that match the Chicxulub impactor. 
In contrast, even after including tidal disruption, the mean timescale for impacts by $D = 7$ km comets is $> 2$ Gyr, in tension with the recency of the Chicxulub impact, as there is only a $\sim 10\%$ probability of such an impact in the last 250 Myr.
Because of the flaws in their interpretation of the literature, the dynamical and geochemical arguments presented by Siraj \& Loeb \citep{Siraj+Loeb2021} do not change the consensus that an asteroid, not a comet, struck the Earth 66 Myr ago.

\bibliography{sample}

\begin{thebibliography}{10}
\urlstyle{rm}
\expandafter\ifx\csname url\endcsname\relax
  \def\url#1{\texttt{#1}}\fi
\expandafter\ifx\csname urlprefix\endcsname\relax\def\urlprefix{URL }\fi
\expandafter\ifx\csname doiprefix\endcsname\relax\def\doiprefix{DOI: }\fi
\providecommand{\bibinfo}[2]{#2}
\providecommand{\eprint}[2][]{\url{#2}}

\bibitem{Siraj+Loeb2021}
\bibinfo{author}{{Siraj}, A.} \& \bibinfo{author}{{Loeb}, A.}
\newblock \bibinfo{journal}{\bibinfo{title}{{Breakup of a long-period comet as
  the origin of the dinosaur extinction}}}.
\newblock {\emph{\JournalTitle{Scientific Reports}}}
  \textbf{\bibinfo{volume}{11}}, \bibinfo{pages}{3803},
  \doiprefix\url{10.1038/s41598-021-82320-2} (\bibinfo{year}{2021}).
\newblock \eprint{2102.06785}.

\bibitem{Alvarez+1980}
\bibinfo{author}{{Alvarez}, L.~W.}, \bibinfo{author}{{Alvarez}, W.},
  \bibinfo{author}{{Asaro}, F.} \& \bibinfo{author}{{Michel}, H.~V.}
\newblock \bibinfo{journal}{\bibinfo{title}{{Extraterrestrial Cause for the
  Cretaceous-Tertiary Extinction}}}.
\newblock {\emph{\JournalTitle{Science}}} \textbf{\bibinfo{volume}{208}},
  \bibinfo{pages}{1095--1108}, \doiprefix\url{10.1126/science.208.4448.1095}
  (\bibinfo{year}{1980}).

\bibitem{Rampino+Stothers1984}
\bibinfo{author}{{Rampino}, M.~R.} \& \bibinfo{author}{{Stothers}, R.~B.}
\newblock \bibinfo{journal}{\bibinfo{title}{{Terrestrial mass extinctions,
  cometary impacts and the Sun's motion perpendicular to the galactic plane}}}.
\newblock {\emph{\JournalTitle{Nature}}} \textbf{\bibinfo{volume}{308}},
  \bibinfo{pages}{709--712}, \doiprefix\url{10.1038/308709a0}
  (\bibinfo{year}{1984}).

\bibitem{Artemieva+Morgan2009}
\bibinfo{author}{{Artemieva}, N.} \& \bibinfo{author}{{Morgan}, J.}
\newblock \bibinfo{journal}{\bibinfo{title}{{Modeling the formation of the K-Pg
  boundary layer}}}.
\newblock {\emph{\JournalTitle{Icarus}}} \textbf{\bibinfo{volume}{201}},
  \bibinfo{pages}{768--780}, \doiprefix\url{10.1016/j.icarus.2009.01.021}
  (\bibinfo{year}{2009}).

\bibitem{Brittan1997}
\bibinfo{author}{{Brittan}, J.}
\newblock \bibinfo{journal}{\bibinfo{title}{{Iridium at the K/T boundary - the
  impact strikes back.}}}
\newblock {\emph{\JournalTitle{Astronomy and Geophysics}}}
  \textbf{\bibinfo{volume}{38}}, \bibinfo{pages}{19--21}
  (\bibinfo{year}{1997}).

\bibitem{Ivanov2005}
\bibinfo{author}{{Ivanov}, B.~A.}
\newblock \bibinfo{journal}{\bibinfo{title}{{Numerical Modeling of the Largest
  Terrestrial Meteorite Craters}}}.
\newblock {\emph{\JournalTitle{Solar System Research}}}
  \textbf{\bibinfo{volume}{39}}, \bibinfo{pages}{381--409},
  \doiprefix\url{10.1007/s11208-005-0051-0} (\bibinfo{year}{2005}).

\bibitem{Bottke+2007}
\bibinfo{author}{{Bottke}, W.~F.}, \bibinfo{author}{{Vokrouhlick{\'y}}, D.} \&
  \bibinfo{author}{{Nesvorn{\'y}}, D.}
\newblock \bibinfo{journal}{\bibinfo{title}{{An asteroid breakup 160Myr ago as
  the probable source of the K/T impactor}}}.
\newblock {\emph{\JournalTitle{Nature}}} \textbf{\bibinfo{volume}{449}},
  \bibinfo{pages}{48--53}, \doiprefix\url{10.1038/nature06070}
  (\bibinfo{year}{2007}).

\bibitem{Morbidelli+2020}
\bibinfo{author}{{Morbidelli}, A.} \emph{et~al.}
\newblock \bibinfo{journal}{\bibinfo{title}{{Debiased albedo distribution for
  Near Earth Objects}}}.
\newblock {\emph{\JournalTitle{Icarus}}} \textbf{\bibinfo{volume}{340}},
  \bibinfo{pages}{113631}, \doiprefix\url{10.1016/j.icarus.2020.113631}
  (\bibinfo{year}{2020}).
\newblock \eprint{2001.03550}.

\bibitem{Kyte1998}
\bibinfo{author}{{Kyte}, F.~T.}
\newblock \bibinfo{journal}{\bibinfo{title}{{A meteorite from the
  Cretaceous/Tertiary boundary}}}.
\newblock {\emph{\JournalTitle{Nature}}} \textbf{\bibinfo{volume}{396}},
  \bibinfo{pages}{237--239}, \doiprefix\url{10.1038/24322}
  (\bibinfo{year}{1998}).

\bibitem{Trinquier+2006}
\bibinfo{author}{{Trinquier}, A.}, \bibinfo{author}{{Birck}, J.-L.} \&
  \bibinfo{author}{{Jean All{\`e}gre}, C.}
\newblock \bibinfo{journal}{\bibinfo{title}{{The nature of the KT impactor. A
  $^{54}$Cr reappraisal}}}.
\newblock {\emph{\JournalTitle{Earth and Planetary Science Letters}}}
  \textbf{\bibinfo{volume}{241}}, \bibinfo{pages}{780--788},
  \doiprefix\url{10.1016/j.epsl.2005.11.006} (\bibinfo{year}{2006}).

\bibitem{Goderis+2013}
\bibinfo{author}{{Goderis}, S.} \emph{et~al.}
\newblock \bibinfo{journal}{\bibinfo{title}{{Reevaluation of siderophile
  element abundances and ratios across the Cretaceous-Paleogene (K-Pg)
  boundary: Implications for the nature of the projectile}}}.
\newblock {\emph{\JournalTitle{Geochimica et Cosmochimica Acta}}}
  \textbf{\bibinfo{volume}{120}}, \bibinfo{pages}{417--446},
  \doiprefix\url{10.1016/j.gca.2013.06.010} (\bibinfo{year}{2013}).

\bibitem{Campins+Swindle1998}
\bibinfo{author}{{Campins}, H.} \& \bibinfo{author}{{Swindle}, T.~D.}
\newblock \bibinfo{journal}{\bibinfo{title}{{Expected characteristics of
  cometary meteorites}}}.
\newblock {\emph{\JournalTitle{Meteoritics and Planetary Science}}}
  \textbf{\bibinfo{volume}{33}}, \bibinfo{pages}{1201--1211},
  \doiprefix\url{10.1111/j.1945-5100.1998.tb01305.x} (\bibinfo{year}{1998}).

\bibitem{Gounelle+2006}
\bibinfo{author}{{Gounelle}, M.}, \bibinfo{author}{{Spurn{\'y}}, P.} \&
  \bibinfo{author}{{Bland}, P.~A.}
\newblock \bibinfo{journal}{\bibinfo{title}{{The orbit and atmospheric
  trajectory of the Orgueil meteorite from historical records}}}.
\newblock {\emph{\JournalTitle{Meteoritics and Planetary Science}}}
  \textbf{\bibinfo{volume}{41}}, \bibinfo{pages}{135--150},
  \doiprefix\url{10.1111/j.1945-5100.2006.tb00198.x} (\bibinfo{year}{2006}).

\bibitem{Hahn+Rettig1998}
\bibinfo{author}{{Hahn}, J.~M.} \& \bibinfo{author}{{Rettig}, T.~W.}
\newblock \bibinfo{journal}{\bibinfo{title}{{Tidal disruption of strengthless
  rubble piles - a dimensional analysis}}}.
\newblock {\emph{\JournalTitle{Planetary and Space Science}}}
  \textbf{\bibinfo{volume}{46}}, \bibinfo{pages}{1677--1682},
  \doiprefix\url{10.1016/S0032-0633(98)00055-5} (\bibinfo{year}{1998}).

\bibitem{Bailey+1992}
\bibinfo{author}{{Bailey}, M.~E.}, \bibinfo{author}{{Chambers}, J.~E.} \&
  \bibinfo{author}{{Hahn}, G.}
\newblock \bibinfo{journal}{\bibinfo{title}{{Origin of sungrazers - A frequent
  cometary end-state.}}}
\newblock {\emph{\JournalTitle{Astronomy and Astrophysics}}}
  \textbf{\bibinfo{volume}{257}}, \bibinfo{pages}{315--322}
  (\bibinfo{year}{1992}).

\end{thebibliography}

%1. Alvarez et al. 1980
%2. Artemieva and Morgan 2009
%3. Bailey et al. 1992
%4. Bottke et al. 2007
%5. Brittan 1997
%6. Campins and Swindle 1998
%7. Davis et al. 1984. 
%8. Goderis et al. 2013
%9. Gounelle et al. 2006
%10. Hahn and Rettig 1998
%11. Morbidelli et al. 2020
%12. Kyte 1998
%13. Rampino and Stothers 1984
%14. Siraj and Loeb 2021
%15. Trinquier et al. 2006

%\noindent LaTeX formats citations and references automatically using the bibliography records in your .bib file, which you can edit via the project menu. Use the cite command for an inline citation, e.g.  \cite{Hao:gidmaps:2014}.

%For data citations of datasets uploaded to e.g. \emph{figshare}, please use the \verb|howpublished| option in the bib entry to specify the platform and the link, as in the \verb|Hao:gidmaps:2014| example in the sample bibliography file.

%\section*{Acknowledgements (not compulsory)}
%
%Acknowledgements should be brief, and should not include thanks to anonymous referees and editors, or effusive comments. Grant or contribution numbers may be acknowledged.

\section*{Author contributions statement}

S.D. led the writing of this manuscript. A.J., J.N., and A.A. contributed ideas. All authors reviewed the manuscript.

\section*{Additional information}

The authors declare no competing interests.

%To include, in this order: \textbf{Accession codes} (where applicable); %\textbf{Competing interests} (mandatory statement). 

%The corresponding author is responsible for submitting a \href{http://www.nature.com/srep/policies/index.html#competing}{competing interests statement} on behalf of all authors of the paper. This statement must be included in the submitted article file.

%\begin{figure}[ht]
%\centering
%\includegraphics[width=\linewidth]{stream}
%\caption{Legend (350 words max). Example legend text.}
%\label{fig:stream}
%\end{figure}

%\begin{table}[ht]
%\centering
%\begin{tabular}{|l|l|l|}
%\hline
%Condition & n & p \\
%\hline
%A & 5 & 0.1 \\
%\hline
%B & 10 & 0.01 \\
%\hline
%\end{tabular}
%\caption{\label{tab:example}Legend (350 words max). Example legend text.}
%\end{table}

%Figures and tables can be referenced in LaTeX using the ref command, e.g. Figure \ref{fig:stream} and Table \ref{tab:example}.

\end{document}